\documentclass[10pt,a4paper,english]{article}
\usepackage[T1]{fontenc}
\usepackage[latin9]{inputenc}
\usepackage{color}
\usepackage{amsmath}
\usepackage{amssymb}
\usepackage{graphicx}
\makeatletter
\usepackage{amsfonts}
\newcommand{\Mn}{M_n(\mathbb{C})}

\newcommand{\ot}{{\,\otimes\,}}
\newcommand{{\Cd}}{{\mathbb{C}^d}}
\newcommand{{\Cn}}{{\mathbb{C}^n}}
\newcommand{{\cp}}{{completely positive\ }}
\newcommand{{\ew}}{{entanglement witness\ }}
\newcommand{{\ews}}{{entanglement witnesses\ }}
\newcommand{{\spa}}{{structural physical approximation\ }}

\def\oper{{\mathchoice{\rm 1\mskip-4mu l}{\rm 1\mskip-4mu l}%
{\rm 1\mskip-4.5mu l}{\rm 1\mskip-5mu l}}}\def\<{\langle}\def\>{\rangle}
\newtheorem{theorem}{Theorem}
\newtheorem{corollary}{Corollary}
\newtheorem{proposition}{Proposition}
\newtheorem{DEF}{Definition}
\newtheorem{remark}{Remark}

\newtheorem{lemma}{Lemma}
\newtheorem{conjecture}{Conjecture}

\begin{document}

\title{\textbf{Optimal entanglement witnesses from generalized reduction and Robertson maps}}

\author{Dariusz Chru\'{s}ci\'{n}ski and Justyna Pytel\\
 Institute of Physics, Nicolaus Copernicus University,\\
 Grudzi\c{a}dzka 5/7, 87--100 Toru\'{n}, Poland}


\maketitle

\begin{abstract}
We provide a generalization of the reduction and Robertson positive
maps in matrix algebras. They give rise to a new class of optimal
entanglement witnesses. Their  structural physical approximation is
analyzed. As a byproduct we provide  new examples of PPT (Positive
Partial Transpose) entangled states.
\end{abstract}


\section{Introduction}

The interest on  quantum entanglement has dramatically increased
during the last two decades due to the emerging field of quantum
information theory \cite{QIT}. It turns out that quantum
entanglement may be used as basic resources in quantum information
processing and communication. The prominent examples are quantum
cryptography, quantum teleportation, quantum error correction codes
and quantum computation.

Since the quantum entanglement is the basic resource for the new
quantum information technologies it is therefore clear that there is
a considerable interest in efficient theoretical and experimental
methods of entanglement detection (see \cite{HHHH}  and \cite{Guhne}
for the review).

Let us recall that a quantum state represented by the density
operator in $\mathcal{H}_A \ot \mathcal{H}_B$ is separable if and
only if it can be represented as a convex combination of product
states
\begin{equation}\label{}
    \rho = \sum_\alpha p_\alpha \rho^{(A)}_\alpha \ot
    \rho^{(B)}_\alpha\ ,
\end{equation}
where $p_\alpha$ denotes a probability distribution whereas
$\rho^{(A)}_\alpha$ and $\rho^{(B)}_\alpha$ are density operators of
A and B subsystem, respectively. It is clear that separable states
define a convex subset in the space of all density operators in
$\mathcal{H}_A \ot \mathcal{H}_B$ and states which are not separable
are called entangled.  The most general approach to characterize
quantum entanglement uses a notion of an entanglement witness (EW)
\cite{EW1,EW2}. A Hermitian operator $W$ defined on a tensor product
$\mathcal{H}_A \ot \mathcal{H}_B$ is called an entanglement witness if and only if:
1) $\mbox{Tr}(W\sigma_{\rm sep})\geq 0$ for all separable states
$\sigma_{\rm sep}$, and 2) there exists an entangled state $\rho$
such that $\mbox{Tr}(W\rho)<0$ (one says that $\rho$ is detected by
$W$).

It turns out that a state is entangled if and only if it is detected
by some EW \cite{EW1}. In recent years there was a considerable
effort in constructing and analyzing the structure of EWs
\cite{Terhal2}--\cite{Justyna-2}. In particular several procedures
for optimizing EWs for arbitrary states were proposed
\cite{O,O1,O2,O3}. Each entangled state $\rho$ may be detected by a
specific choice of $W$. It is therefore clear that each EW provides
a new separability test and it may be interpreted as a new type of
Bell inequality \cite{Lew3}. There is, however, no general procedure
for constructing  EWs.

In this paper we provide a new class of EWs. It is well known (see
the next section for all details) that each EW is uniquely related
to a linear positive map $\Lambda : \mathcal{B}(\mathcal{H}_A)
\rightarrow \mathcal{B}(\mathcal{H}_B)$. We provide new classes of
linear positive maps by constructing generalization of well known
maps, namely reduction map and Robertson map. It is shown that
generalized maps and corresponding witnesses are optimal, that is,
they detect quantum entanglement in an `optimal way' (see next
section for the precise definition). Optimal EWs are of primary
importance since to perform complete classification of quantum
states of a bipartite system it is enough to use only optimal EWs.
Finally, we discuss how these maps are related to the idea of
physical structural approximation (SPA) \cite{SPA1,SPA2,SPA3}. It is
shown that there is a strong evidence that these EWs support the
conjecture  \cite{SPA3} (see also  \cite{SPA4}) that physical
structural approximation to optimal positive map gives rise to an
entanglement breaking channel.

The paper  is organized as follows: we recall in
Section~\ref{NOTATION} basic facts about linear positive maps and
entanglement witnesses. Section \ref{RED} discusses generalization
of the reduction map whereas Section \ref{ROB} discusses
generalization of the Robertson map. We show that these maps and the
corresponding entanglement witnesses are optimal. Final conclusions
are collected in the last Section.

\section{Preliminaries and notation}  \label{NOTATION}

In this paper we consider finite dimensional complex Hilbert spaces.
Let $\Mn$ denote an algebra (actually, a $\mathbb{C}^{*}$-algebra)
of $n\times n$ complex matrices. A linear map $\Lambda : \Mn
\rightarrow M_m(\mathbb{C})$ is called to be positive if it maps
positive elements from $\Mn$ into positive elements in
$M_m(\mathbb{C})$. It means that for any vectors $|x\> \in
\mathbb{C}^n$ and $|y\> \in \mathbb{C}^m$ one has
\begin{equation}\label{}
    {\rm Tr}(P_y \Lambda(P_x)) \geq 0\ ,
\end{equation}
where $P_x = |x\>\<x|$ and $P_y= |y\>\<y|$. Equivalently, $ \< y |
\Lambda(|x\>\<x|)|y\> \geq 0\,.$ Note, that the above condition is
in general very hard to check since it does not reduce to any
spectral condition.  Unfortunately, in spite of the considerable
effort, the structure of positive maps is rather poorly understood
\cite{Stormer1}--\cite{Woronowicz2} (see also the monograph by
Paulsen \cite{Paulsen}). For some recent works see
\cite{Skowronek-1,Skowronek-2,CMP,kule,Justyna-1,Justyna-2,Gniewko}
and for a review paper see \cite{OSID-W}. Positive maps play an
important role both in physics and mathematics providing
generalization of $*$-homomorphisms, Jordan homomorphisms and
conditional expectations. Normalized positive maps define affine
mappings between sets of states of $\mathbb{C}^*$-algebras. A
positive linear map $\Lambda$ is $k$-positive if the map
\begin{equation}\label{}
    \oper_k \ot \Lambda : M_k(\Mn) \longrightarrow
M_k(M_m(\mathbb{C})) ,
\end{equation}
is positive ($M_k(\mathcal{A})$ denotes a set of $k\times k$ complex
matrices with entries from the $\mathbb{C}^*$-algebra
$\mathcal{A}$). Clearly, a $k$-positive map is $l$-positive for all
$l < k$. A map which is $k$-positive for all $k$ is called
completely positive. Actually, in the finite dimensional case we
consider in this paper $\Lambda$ is \cp if and only if it is $k$
positive with $k = \min\{n,m\}$ \cite{Choi1}.

Let $\{e_1,\ldots,e_n\}$ be a fixed orthonormal basis in
$\mathbb{C}^n$. Denote by $e_{ij} := |e_i\>\<e_j|$ an orthonormal
basis in $\Mn$. Let ${\rm T} : \Mn \longrightarrow \Mn$ denotes
transposition map with respect to the fixed basis $\{e_i\}$, that is
${\rm T}(e_{ij}) = e_{ji}$. Evidently, `T' defines linear positive
map. Now, a positive map $\Lambda$ is called decomposable if and
only if
\begin{equation}\label{}
    \Lambda = \Lambda_1 +  \Lambda_2 \circ {\rm T}\ ,
\end{equation}
where $\Lambda_1$ and $\Lambda_2$ are \cp. Maps which are not
decomposable are called indecomposable (or nondecomposable).

Using Choi-Jamio{\l}kowski \cite{Choi1,Jam} isomorphism each
positive map $\Lambda$ gives rise to entanglement witness $W$
\begin{equation}\label{CJ}
    W = (\oper_n \ot \Lambda)P^+_n\ ,
\end{equation}
where $P^+_n$ denotes maximally entangled state in $\mathbb{C}^{n}
\ot \mathbb{C}^{n}$ and $\oper_n$ denotes an identity map acting on
$\Mn$. Using fixed basis $\{e_i\}$ one has
\begin{equation}\label{}
    W = \frac 1n \sum_{i,j=1}^n e_{ij} \ot \Lambda(e_{ij})\ .
\end{equation}
An entanglement witness $W$ is called (in)decomposable if the
corresponding positive map $\Lambda$ is (in)decomposable. Hence, any
decomposable entanglement witness may be represented as follows
\begin{equation}\label{}
    W = Q_1 + Q_2^\Gamma\ ,
\end{equation}
where $Q_1,Q_2 \geq 0$, and $A^\Gamma := (\oper_n \ot {\rm T})A$
denotes partial transposition of $A$.  Let us observe that the
positivity of $\Lambda$ implies that $W$ satisfies
\begin{equation}\label{W-xy}
    \< x \ot y |W|x \ot y\> \geq 0\ ,
\end{equation}
for any vectors $|x\> \in \mathbb{C}^n$ and $|y\> \in \mathbb{C}^m$.
Hermitian operators satisfying (\ref{W-xy}) are often called
block-positive. Note, that if $\Lambda$ is \cp then the
corresponding $W$ is not only block-positive but even positive.

Let us recall that entanglement witnesses play a key role in the
theory of entanglement. A density operator $\rho$ living in
$\mathbb{C}^n \ot \mathbb{C}^m$ is entangled if and only if there exists an entanglement witness
$W$ such that
\begin{equation}\label{}
    {\rm Tr}(W \rho) < 0\ .
\end{equation}
One says that $\rho$ is detected by $W$. Recall, that a state
represented by a density operator $\rho$ is PPT (Positive Partial
Transpose) if $\rho^\Gamma \geq 0$. One has \cite{Stormer1,O}

\begin{proposition}
$W$ is an indecomposable entanglement witness if and only if there exists a PPT state
$\rho$ detected by $W$. Equivalently, a PPT state $\rho$ is
entangled if and only if there exists an indecomposable entanglement witness which
detects $\rho$.
\end{proposition}
Let $\mathcal{D}$ be a subset of density operators of a composite
quantum system living in $\mathbb{C}^n \ot \mathbb{C}^m$ detected by
a given entanglement witness $W$, i.e. $\mathcal{D} = \{ \rho\ | \ {\rm Tr}(W \rho) <
0 \}$. Given two entanglement witnesses $W_1 $ and $W_2$ one says that $W_2$ is finer
than $W_1$ if $\mathcal{D}_1 \subset \mathcal{D}_2$, that is, all
states detected by $W_1$ are also detected by $W_2$. A witness $W$
is optimal if there is no other entanglement witness which is finer than $W$. It
means that $W$ detects quantum entanglement in the `optimal way'. It
is clear that the knowledge of optimal entanglement witnesses is crucial to classify
quantum states of composite systems. One proves \cite{O} the
following

\begin{proposition}
$W$ is an optimal entanglement witness if and only if $W- Q$ is no longer entanglement witness for
arbitrary positive operator $Q$.
\end{proposition}
Authors of Ref.  \cite{O} formulated the following criterion for the
optimality of $W$.

\begin{proposition}  \label{PRO-Lew}
If the set of product vectors $x \ot y \in \mathbb{C}^n \ot
\mathbb{C}^m$ satisfying
\begin{equation}\label{W-opt}
    \< x \ot y|W|x \ot y\>=0\ ,
\end{equation}
span the total Hilbert space $\mathbb{C}^n \ot \mathbb{C}^m$, then
$W$ is optimal.
\end{proposition}
It should be stressed that the converse theorem is not true, i.e.
the existence of product vectors which span $\mathbb{C}^n \ot
\mathbb{C}^m$ and satisfy (\ref{W-opt}) is not necessary for the
optimality of $W$. A well know example is provided by the entanglement witness
corresponding to the celebrated Choi indecomposable map \cite{Choi1}
which is known to be optimal but does not provide the corresponding
collection of $|x \ot y\>$.

Finally, let us comment on an interesting conjecture proposed in
\cite{SPA3}: let $W$ be a normalized entanglement witness, i.e.
${\rm Tr}\,W=1$. An operator $\widetilde{W}(p)$ defined by
\begin{equation}
\widetilde{W}(p)=\frac{1-p}{n^{2}}\,\mathbb{I}_{n}\ot\mathbb{I}_{n}+pW
\,\label{}
\end{equation}
is called \spa (SPA) of $W$ if $\widetilde{W}(p) \geq 0$. Now, let
$p_*$ be a maximal $p$ for which $\widetilde{W}(p)$ defines SPA for
$W$, that is, $\widetilde{W}(p) \geq 0$ for $p\in [0,p_*]$.
\begin{conjecture}
If $W$ is an optimal entanglement witness, then $\widetilde{W}(p_*)$ defines a separable state.
\end{conjecture}
It should be clear that SPA can be equivalently defined for a positive map
$\Lambda : \Mn \longrightarrow \Mn $. Let us recall \cite{EB}

\begin{DEF}
A completely positive map $\Lambda : \Mn \longrightarrow M_n(\mathbb{C})$
is entanglement breaking if and only if $(\oper_n \ot \Lambda)\rho$ defines a  separable
state  for any $\rho$ living  in $\Cn \ot \Cn$.
\end{DEF}
Interestingly, any entanglement breaking quantum channel (trace preserving completely positive map)
can be represented in the Holevo form \cite{Holevo}
\begin{equation}\label{}
    \Lambda(\rho) = \sum_i R_i {\rm Tr}(F_i \rho)\ ,
\end{equation}
where $R_i$ are density operators in $\mathbb{C}^m$ and $F_i$ are positive operators in $\Cn$
satisfying $\sum_i F_i = \mathbb{I}_n$, i.e. a set $\{F_i\}$ defines a generalized quantum measurement.
Now,  a positive map $\Lambda$ is optimal if $\Lambda - \Phi$, with $\Phi$ being a completely positive map,
 is no longer positive. A positive map
\begin{equation}\label{}
    \widetilde{\Lambda}(p) = (1-p) \oper_n + p \Lambda\ ,
\end{equation}
defines  a SPA for  $\Lambda$ if $\widetilde{\Lambda}(p)$ is
completely positive. The above conjecture may be equivalently
formulated as follows: if $\Lambda$ is an optimal positive map, then
$\widetilde{\Lambda}(p_*)$ is entanglement breaking. One proves
\cite{Justyna-1} the following

\begin{theorem} \label{TH-nasze} Let $\Lambda:\Mn\rightarrow\Mn$
be a unital map (i.e. $\Lambda(\mathbb{I}_n) = \mathbb{I}_n$) that
detects all entangled isotropic states. Then SPA of $\Lambda$ is an
entanglement breaking map.
\end{theorem}
Let $\widetilde{W}(p)$
be SPA of $W$  and let $\lambda_{{\rm min}}$ be the smallest eigenvalue of $W$.
One easily finds
\begin{equation}
p_*=\frac{1}{1+|\lambda_{{\rm min}}|n^{2}}\ .\label{}
\end{equation}
Now, it follows from Theorem~\ref{TH-nasze} that $p_* =
\frac{1}{n+1}$ and hence

\begin{corollary} \label{l-min} If $\Lambda:\Mn\rightarrow\Mn$
is a unital map, and the smallest eigenvalue of the corresponding
entanglement witness $W$ satisfies \begin{equation}
\lambda_{{\rm min}}\leq-\frac{1}{n}\ ,\label{}
\end{equation}
 then SPA of ${W}$ defines a separable state.
\end{corollary}

Conjecture 1 is supported by several examples (see  \cite{SPA3}
and \cite{Justyna-1,Justyna-2}). The present paper provides another
family of examples supporting above conjecture.

\section{New optimal EWs out of the reduction map}   \label{RED}

\subsection{Reduction map in $\Mn$}

Let us start with an elementary positive map in $\Mn$ called
reduction map
\begin{equation}
R_{n}(X)=\frac{1}{n-1}\Big[\mathbb{I}_{n}\mathrm{Tr}X-X\Big]\
,\label{Rn}
\end{equation}
for $X\in\Mn$. Positivity of $R_{n}$ follows from the fact that
$R_{n}$ maps rank-1 projectors into projectors. Indeed, for
$X=|\psi\>\<\psi|$ with $\<\psi|\psi\>=1$, one has
\begin{equation}
R_{n}(|\psi\>\<\psi|)=\frac{1}{n-1}\Big[\mathbb{I}_{n}-|\psi\>\<\psi|\Big]\
,\label{}\end{equation}
 which is evidently positive, since $\mathbb{I}_{n}-|\psi\>\<\psi|$
is a projector (of rank `$n-1$') onto the $(n-1)$--dimensional
hyperplane orthogonal to $|\psi\>$. The corresponding entanglement
witness is given by
\begin{equation}
W=\frac{1}{n-1}\Big(\frac{1}{n}\mathbb{I}_{n}\ot\mathbb{I}_{n}-P_{n}^{+}\Big)\
.\label{W-}
\end{equation}
One has for the partial transposition \begin{equation} (\oper\ot{\rm
T})W = \frac{1}{n(n-1)}\sum_{i<j}P_{ij}\ ,\label{dec-}\end{equation}
where
\begin{equation} P_{ij}=|\psi_{ij}\>\<\psi_{ij}|\
 ,\label{}\end{equation}
 with \begin{equation}
|\psi_{ij}\>= e_i \ot e_j - e_j \ot e_i\ ,\label{}\end{equation}
 which shows that $(\oper_n \ot{\rm T})W \geq0$ and hence $W$
defines a decomposable EW. Equivalently, it shows that the map
$R_{n}\circ{\rm T}$ is completely positive, i.e. it defines a
legitimate quantum channel. Note, that decomposition (\ref{dec-})
proves that $W$ is not extremal, since it decomposes into a convex
combination of extremal witnesses $P_{ij}^{\Gamma}$ (it is extremal
for $n=2$ only, due to $W=\frac{1}{2}P_{12}^{-\,\Gamma}$).
Interestingly, being not extremal it is still optimal.

\begin{proposition} $W$ is an optimal EW. \end{proposition}
{\em Proof}: to show that $W$ is optimal we use Proposition
\ref{PRO-Lew}.  Let us introduce the following set of vectors in
$\Cn \ot \Cn$:
\begin{eqnarray*} f_{kl}=(e_k + e_l)\ot (e_k + e_l) \ ,\ \ \ g_{kl}=
(e_k + i e_l)\ot (e_k -i e_l) ,\end{eqnarray*}
 for each $1\leq k<l\leq n$. It is easy to check that $n^{2}$ vectors
 $\,\{\,e_k \ot e_k \,,f_{kl}\,,g_{kl}\,\}$ are linearly
independent and hence they do span $\mathbb{C}^{n}\ot\mathbb{C}^{n}$.
Direct calculation shows that
\begin{equation}
\< f_{kl} |W| f_{kl}\>=0\ , \ \ \< g_{kl} |W| g_{kl}\>=0\ , \ \ \< e_k \ot e_k |W| e_k \ot e_k\>=0\ , \
\label{}\end{equation}
 which ends the proof. \hfill{}$\Box$

Finally, the reduction map $R_{n}$ supports recent conjecture
\cite{SPA3}, that is, one has the following

\begin{proposition} \label{SPA-Rn} The structural physical approximation
of $R_{n}$ is an entanglement breaking map. \end{proposition}

Let as observe that the smallest eigenvalue of $W$ is given by
$\lambda_{{\rm min}}=-1/n$, and hence, due to Corollary~\ref{l-min},
SPA of $W$ is separable. Actually, the above proposition was already
proved in \cite{SPA3}.

\subsection{Generalized reduction map}

Let us observe that taking the orthonormal basis $e_{ij}$ in
$M_{n}(\mathbb{C})$ the reduction map $R_{n}$ may be defined as
follows \begin{eqnarray}
R_{n}(e_{ii}) & = & \frac{1}{n-1}\,(\mathbb{I}_{n}-e_{ii})\ ,\label{}\\
R_{n}(e_{ij}) & = & -\frac{1}{n-1}\, e_{ij}\ ,\ \ \ i\neq j\ .\end{eqnarray}
 Let us take $n(n-1)/2$ complex numbers $z_{ij}$ $(i<j)$ satisfying
$|z_{ij}|\leq1$ and denote by $\mathbf{z}$ the collection
$\{z_{12},\ldots,z_{n-1,n}\}$. Finally, let us define a map
\begin{equation} R_{n}^{(\mathbf{z})}\ :\ \Mn\longrightarrow\ \Mn\
,\label{Rn-z}\end{equation}
 by \begin{eqnarray}
R_{n}^{({\bf z})}(e_{ii}) & = & \frac{1}{n-1}\,(\mathbb{I}_{n}-e_{ii})\ ,\label{}\\
R_{n}^{({\bf z})}(e_{ij}) & = & -\frac{z_{ij}}{n-1}\, e_{ij}\ ,\ \ \
i<j\ ,\end{eqnarray} and $z_{ij} = \overline{z_{ji}}\,$ for $i>j$.
It is clear that for $z_{ij}=1$ one reconstructs the original
reduction map $R_{n}$.

\begin{proposition} $R_{n}^{({\bf z})}$ defines a positive decomposable
map. \end{proposition}

\noindent Proof: let us observe that the corresponding entanglement
witness $W_{n}^{({\bf z})}$ has the following form \begin{equation}
W^{({\bf z})}=\frac{1}{n(n-1)}\sum_{i,j=1}^{n}e_{ij}\ot W_{ij}^{({\bf z})}\ ,\label{}\end{equation}
 where \begin{equation}
W_{ii}^{({\bf z})}=\mathbb{I}_{n}-e_{ii}\ ,\ \ \ \ W_{ij}^{({\bf
z})}=-z_{ij}e_{ij}\ \ \ (i<j)\ .\label{}\end{equation}
 To complete the proof observe that $(\oper_n \ot{\rm T})W^{({\bf z})}$
is a positive operator. Indeed, one has \begin{equation} (\oper_n
\ot{\rm T})W^{({\bf
z})}=\frac{1}{n(n-1)}\sum_{i<j}P_{ij}^{(\mathbf{z})}\
,\label{}\end{equation}
 where the operators $P_{ij}^{(\mathbf{z})}$ are defined by \begin{equation}
P_{ij}^{(\mathbf{z})}=e_{ii}\ot e_{jj}+e_{jj}\ot e_{ii}-z_{ij}e_{ij}\ot
e_{ji}-\overline{z_{ij}}\, e_{ji}\ot e_{ij}\,\label{}\end{equation}
 and hence they are positive for $|z_{ij}|\leq1$. It shows that $(\oper_n\ot{\rm T})W^{({\bf z})}\geq0$
and hence $W^{({\bf z})}$ is a decomposable entanglement witness.
\hfill{}$\Box$

 Note, that if at least one $z_{ij}\neq0$, then the
map $R_{n}^{({\bf z})}$ is not completely positive. Indeed, the
following principal submatrix of $W^{(\mathbf{z})}$
$$   \left( \begin{array}{cc} 0 & z_{ij} \\ \overline{z_{ij}}  & 0
\end{array} \right)\ , $$
is not positive definite and hence $W^{(\mathbf{z})} \ngeq 0$. If
$|z_{ij}|=1$, then
$P_{ij}^{(\mathbf{z})}=|\psi_{ij}^{(\mathbf{z})}\>\<\psi_{ij}^{(\mathbf{z})}|$,
with \begin{equation} |\psi_{ij}^{(\mathbf{z})}\>=e_{i}\ot e_{j}-
\overline{z_{ij}}\, e_{j}\ot e_{i}\ .\label{}\end{equation}

\begin{proposition} \label{R-opt}  The positive map $R_{n}^{({\bf z})}$ is optimal
if and only if $|z_{ij}|=1$ for all $i\neq j$. \end{proposition}

\noindent Proof: the condition $|z_{ij}|=1$ is necessary for
optimality. Indeed, suppose for example that $|z_{kl}|<1$ for some
pair $k<l$. Then \begin{equation} (\oper_n\ot{\rm T})W^{({\bf
z})}-\frac{1}{n(n-1)}Q_{kl}^{(\mathbf{z})}\ ,\label{}\end{equation}
 where \begin{equation}
Q_{kl}^{(\mathbf{z})}=(1-|z_{kl}|^{2})(e_{kk}\ot e_{ll}+e_{ll}\ot e_{kk})\ ,\label{}\end{equation}
 is still a positive operator, and hence \begin{equation}
W^{({\bf z})}-\frac{1}{n(n-1)}Q_{kl}^{(\mathbf{z})}\ ,\label{}\end{equation}
 defines  decomposable entanglement witness
 (note, that $(\oper_n \ot{\rm T})Q_{kl}^{(\mathbf{z})}=Q_{kl}^{(\mathbf{z})}$).

Suppose now that $|z_{kl}|=1$. To show that $W^{({\bf z})}$ is
optimal we use again the result of Lewenstein et. al. \cite{O}. Let
$z_{kl}=e^{i\alpha_{kl}}$. It is easy to check that the following
vectors
\[
f_{kl}=(e_{k}+e^{-i\alpha_{kl}/2}e_{l})\ot(e_{k}+e^{-i\alpha_{kl}/2}e_{l})\
,\ \ \
g_{kl}=(e_{k}+ie^{-i\alpha_{kl}/2}e_{l})\ot(e_{k}-ie^{-i\alpha_{kl}/2}e_{l})\
,\ \ \ e_{k}\ot e_{k}\ ,\]
 span the entire Hilbert space $\mathbb{C}^{n}\ot\mathbb{C}^{n}$.
Moreover, they satisfy \begin{equation}
\<f_{kl}|W^{({\bf z})}|f_{kl}\>=\<g_{kl}|W^{({\bf z})}|g_{kl}\>=0\label{}\end{equation}
 for $k<l$, and \begin{equation}
\<e_{k}\ot e_{k}|W^{({\bf z})}|e_{k}\ot e_{k}\>=0\ ,\label{}\end{equation}
 for $k=1,\ldots,n$ which proves that $W^{({\bf z})}$ is an optimal
entanglement witness. \hfill{}$\Box$

Finally, consider the structural physical approximation to $W^{({\bf
z})}$
\begin{equation}
\widetilde{W}^{({\bf
z})}(p)=\frac{1-p}{n^{2}}\,\mathbb{I}_{n}\ot\mathbb{I}_{n}+p
W^{({\bf z})}\,\label{}\end{equation}
 and let $\lambda_{{\rm min}}^{({\bf z})}$ be the smallest eigenvalue
of $W^{({\bf z})}$. One has \begin{equation} p_*^{({\bf
z})}=\frac{1}{1+|\lambda_{{\rm min}}^{({\bf z})}|n^{2}}\
.\label{}\end{equation}
 Note, that $\lambda_{{\rm min}}^{({\bf z})}$ is the smallest eigenvalue
to the $n\times n$ Hermitian matrix $Z$ defined by \begin{equation}
Z_{ii}:=0\ ,\ \ \ \ Z_{ij}:=z_{ij}\ (i<j)\ .\label{}\end{equation}
 Note, that if all $z_{ij}=1$ (standard reduction map), then \begin{equation}
\lambda_{{\rm min}}^{({\bf z})}=-\frac{1}{n}\ ,\label{}\end{equation}
 and if all $z_{ij}=-1$, then \begin{equation}
\lambda_{{\rm min}}^{({\bf z})}=-\frac{1}{n(n-1)}\ .\label{p-1}\end{equation}
 For a set of arbitrary $z_{ij}=e^{i\alpha_{ij}}$ the analytic formula
for $\lambda_{{\rm min}}^{({\bf z})}$ is not available. However,
it is clear that in the general case one has \begin{equation}
-\frac{1}{n}\,\leq\,\lambda_{{\rm min}}^{({\bf z})}\,\leq\,-\frac{1}{n(n-1)}\ ,\label{}\end{equation}
 and hence \begin{equation}
\frac{1}{n+1}\,\geq\, p_*^{({\bf z})}\,\geq\,\frac{n-1}{2n+1}\
.\label{}\end{equation}
 We have already shown that for $z_{ij}=1$ the SPA of $W^{({\bf z})}$
defines a separable state (see Proposition \ref{SPA-Rn}).

\begin{proposition} \label{SPA-Rz} The structural physical approximation
$R_{n}^{({\bf z})}(p_*^{({\bf z})})$, with $|z_{ij}|=1$, is an
entanglement breaking map. \end{proposition}

\noindent Proof: one has
\begin{equation}
\widetilde{W}^{({\bf z})}(p_*^{({\bf z})})=\frac{1-p_*^{({\bf
z})}}{n^{2}}\,\mathbb{I}_{n}\ot\mathbb{I}_{n}+p_*^{({\bf z})}\,
W^{({\bf z})}= p_*^{({\bf z})} \Big[ |\lambda^{({\bf z})}_{\rm min}|
\, \mathbb{I}_{n}\ot\mathbb{I}_{n} + W^{({\bf z})} \Big] \ ,\label{}
\end{equation}
and hence to prove the Proposition one has to show that
$$ B^{({\bf z})} = |\lambda^{({\bf z})}_{\rm min}| \, \mathbb{I}_{n}\ot\mathbb{I}_{n} +
W^{({\bf z})}$$ defines a separable positive operator.

\begin{lemma}
A positive operator
\begin{equation}\label{}
    A^{({\bf z})} = \sum_{i,j=1}^n e_{ij} \ot A^{({\bf z})}_{ij} \ ,
\end{equation}
with
\begin{equation}\label{}
    A^{({\bf z})}_{ii} = |\lambda^{({\bf z})}_{\rm min}| \mathbb{I}_n\ ,\
    \ \  A^{({\bf z})}_{ij} = - z_{ij}\, e_{ij} \ , \ \ \ (i< j)\ ,
\end{equation}
is separable.
\end{lemma}

\noindent Proof: consider the following operator living in $\Cn \ot
\Cn$:
\begin{equation}\label{}
    A^{({\bf z})} = \sum_{i,j=1}^n \widetilde{Z}_{ij}\, e_{ij} \ot e_{ij} + |\lambda^{({\bf z})}_{\rm min}|\,
     \sum_{i\neq j} e_{ii} \ot e_{jj} \ ,
\end{equation}
where the $n \times n$ matrix $\widetilde{Z}$ is defined as follows
\begin{equation}\label{}
    \widetilde{Z}_{ii} = |\lambda^{({\bf z})}_{\rm min}| \ , \ \ \ \
    \widetilde{Z}_{ij} = - z_{ij} \ , \ \  (i < j ) \ .
\end{equation}
It is clear that $\widetilde{Z} \geq 0$, and hence $A^{(\mathbf{z})}
\geq 0$. Now, let us define the linear map $
    \Lambda^{(\mathbf{z})}  :  M_{n}(\mathbb{C})
    \longrightarrow M_{n}(\mathbb{C})\,$ defined as follows
\begin{equation}\label{}
\Lambda^{(\mathbf{z})}(X) = \widetilde{Z} \circ X\ ,
\end{equation}
where $\widetilde{Z} \circ X$ denotes the Hadamard product of
matrices $X,\widetilde{Z} \in M_{n}(\mathbb{C})$. Recall, that $[A
\circ B]_{ij} := A_{ij} B_{ij}$. It is well known \cite{Bhatia} that
$\Lambda^{(\mathbf{z})}$ is \cp due to the positivity of the matrix
$\widetilde{Z}$. Observe, that
\begin{equation}\label{AzA0}
 A^{({\bf z})} = (\oper \ot  \Lambda^{({\bf z})}) A_0\ ,
\end{equation}
where
\begin{equation}\label{}
    A_0 = \sum_{i,j=1}^n  e_{ij} \ot e_{ij} +  \sum_{i\neq j} e_{ii} \ot e_{jj} \
    .
\end{equation}
Note, that $A_0 = A^{({\bf z})}$ with $z_{ij}=1$. Now, it is well
known that $A_0$ defines a separable operator and hence due to
(\ref{AzA0}) the operator $A^{({\bf z})}$ is separable as well.
\hfill $\Box$

It is evident that the separability of $B^{({\bf z})}$ follows from
the separability of $A^{({\bf z})}$ which completes the proof of the
Proposition. \hfill $\Box$

\begin{remark} Note, that for $n=2$ all maps $R_2^{({z})}$ with
$|z|=1$ are unitarily equivalent ($z\equiv z_{12}$) \begin{equation}
R_{2}^{({z})}(X)=V^{(z)}R_{2}(X)V^{(z)\,\dagger}\
,\label{}\end{equation}
 with \begin{equation}
V^{(z)}=\left(\begin{array}{cc}
1 & 0\\
0 & \overline{z}\end{array}\right)\ .\label{}\end{equation}
 Clearly, it is not longer true for $n>2$.  \end{remark}

\section{New optimal EWs out of the Robertson map}  \label{ROB}

\subsection{Robertson map in $M_{2k}(\mathbb{C})$}

Robertson provided \cite{Robertson} the following linear map $
\Phi_{4}:M_{4}(\mathbb{C})\longrightarrow M_{4}(\mathbb{C})$
\begin{equation}
\Phi_{4}\left(\begin{array}{c|c}
X_{11} & X_{12}\\
\hline X_{21} & X_{22}\end{array}\right)=\frac{1}{2}\left(\begin{array}{c|c}
\mathbb{I}_{2}\,\mbox{Tr}X_{22} & -[X_{12}+R_{2}(X_{21})]\\
\hline -[X_{21}+R_{2}(X_{12})] & \mathbb{I}_{2}\,\mbox{Tr}X_{11}\end{array}\right)\ ,\label{R4}\end{equation}
 where $X_{kl}\in M_{2}(\mathbb{C})$. It turns out \cite{Robertson}
that $\Phi_{4}$ defines  a unital positive indecomposable map.
Moreover, $\Phi_{4}$ is extremal and hence optimal. Interestingly,
Robertson map supports  the SPA conjecture \cite{SPA3}.

Recently, \cite{Breuer,Hall} (see also discussion in
\cite{Justyna-1,Justyna-2,JPA-2}) Robertson map was generalized to a
linear map $\Phi_{2k}:M_{2k}(\mathbb{C})\longrightarrow
M_{2k}(\mathbb{C})$
\begin{eqnarray}
\Phi_{2k}\left(\begin{array}{c|c|c|c}
X_{11} & X_{12} & \cdots & X_{1k}\\
\hline X_{21} & X_{22} & \cdots & X_{2k}\\
\hline \vdots & \vdots & \ddots & \vdots\\
\hline X_{k1} & X_{k2} & \cdots & X_{kk}\end{array}\right)=\frac{1}{2(k-1)}\left(\begin{array}{c|c|c|c}
A_{1} & -B_{12} & \cdots & -B_{1k}\\
\hline -B_{21} & A_{2} & \cdots & -B_{2k}\\
\hline \vdots & \vdots & \ddots & \vdots\\
\hline -B_{k1} & -B_{k2} & \cdots & A_{k}\end{array}\right)\ ,\label{Phi-2k}\end{eqnarray}
 where \begin{equation}
A_{k}=\mathbb{I}_{2}(\mathrm{Tr}X-\mathrm{Tr}X_{kk})\ ,\label{}\end{equation}
 and \begin{equation}
B_{kl}=X_{kl}-R_{2}(X_{lk})\ .\label{}\end{equation} It was shown
\cite{Breuer} that $\Phi_{2k}$ defines an indecomposable optimal
positive map. Analyzing the spectrum of the corresponding entanglement witness $W  =
(\oper_{2k} \ot \Phi_{2k})P^+_{2k}$ one finds single negative
eigenvalue `$-1/2k$', one strictly positive eigenvalue
`$1/[2k(k-1)]$'  with multiplicity $2k^2 - (k+1)$, and $k(2k+1)$
zero-modes. Therefore, due to the Corollary \ref{l-min} the SPA of
$\Phi_{2k}$ defines an entanglement breaking map and hence supports
conjecture of \cite{SPA3}.

\begin{remark}
Note, that $\Phi_{2k}$ defines a special example of the Breuer-Hall
map \cite{Breuer,Hall}
\begin{equation}\label{BH}
    \Phi^U_{2k}(X) = \frac{1}{2(k-1)} \left( R_{2k}(X)  - U X^{\rm
    T} U^\dagger \right )\ ,
\end{equation}
where $U$ is a unitary antisymmetric $2k \times 2k$ matrix. It
corresponds to
\begin{equation}\label{}
    U = \mathbb{I}_k \ot \sigma_y\ .
\end{equation}
It was shown \cite{Breuer} that for any $U$ the map $\Phi_{2k}^U$ is
indecomposable and optimal. The special form of $\Phi^U_{2k}$
resembling the original Robertson map in $M_4(\mathbb{C})$ was
proposed in \cite{Justyna-1}.
\end{remark}

\subsection{Generalized Robertson map in $M_{2k}(\mathbb{C})$}

In analogy to the reduction map discussed in the previous section we
propose the following generalization of the Robertson map
$\Phi_{2k}$: for any collection of  $k(k-1)/2$ complex numbers
$z_{ij}$, with $i<j$, satisfying $|z_{ij}|\leq1$ we define
$\Phi_{2k}^{({\bf z})}:M_{2k}(\mathbb{C})\longrightarrow
M_{2k}(\mathbb{C})$ by
\begin{eqnarray} \Phi_{2k}^{({\bf z})}\left(\begin{array}{c|c|c|c}
X_{11} & X_{12} & \cdots & X_{1k}\\
\hline X_{21} & X_{22} & \cdots & X_{2k}\\
\hline \vdots & \vdots & \ddots & \vdots\\
\hline X_{k1} & X_{k2} & \cdots & X_{kk}\end{array}\right)=\frac{1}{2(k-1)}\left(\begin{array}{c|c|c|c}
A_{1} & -z_{12}B_{12} & \cdots & -z_{1k}B_{1k}\\
\hline -{\overline{z}_{21}}B_{21} & A_{2} & \cdots & -{z}_{2k}\, B_{2k}\\
\hline \vdots & \vdots & \ddots & \vdots\\
\hline -\overline{z}_{k1}\, B_{k1} & -\overline{z}_{k2}\, B_{k2} & \cdots & A_{k}\end{array}\right)\ .\label{Phi-2k}\end{eqnarray}
 The main result of this section consists in the following

\begin{theorem} \label{MAIN} $\Phi_{2k}^{({\bf z})}$ defines a
positive map. \end{theorem}

\noindent Proof: to prove the positivity of $\Phi_{2k}^{({\bf z})}$
one has to show that for any rank-1 projector
$P_{2k}=|\psi\>\<\psi|$, one has
\begin{equation} \Phi_{2k}^{({\bf z})}(P_{2k})\geq0\
,\label{!}\end{equation}
 where $\psi\in\mathbb{C}^{2k}$ and $\<\psi|\psi\>=1$. Now, any
normalized $|\psi\>\in\mathbb{C}^{2k}$ may be considered as a direct
sum \begin{equation}
|\psi\>=\sqrt{\alpha_{1}}\,|\psi_{1}\>\oplus\ldots\oplus\sqrt{\alpha_{k}}\,|\psi_{k}\>\ ,\label{}\end{equation}
 where $|\psi_{i}\>\in\mathbb{C}^{2}$, such that $\<\psi_{i}|\psi_{i}\>=1$,
and $\alpha_{1},\ldots,\alpha_{k}\geq0$ satisfy normalization condition
\begin{equation}
\alpha_{1}+\ldots+\alpha_{k}=1\ .\label{}\end{equation}
 Using such representation the projector $P_{2k}=|\psi\>\<\psi|$ has
the following form \begin{equation}
P_{2k}=\left(\begin{array}{c|c|c|c}
\alpha_{1}|\psi_{1}\rangle\langle\psi_{1}| & \sqrt{\alpha_{1}\alpha_{2}}|\psi_{1}\rangle\langle\psi_{2}| & \cdots & \sqrt{\alpha_{1}\alpha_{k}}|\psi_{1}\rangle\langle\psi_{k}|\\
\hline \sqrt{\alpha_{2}\alpha_{1}}|\psi_{2}\rangle\langle\psi_{1}| & \alpha_{2}|\psi_{2}\rangle\langle\psi_{2}| & \cdots & \sqrt{\alpha_{2}\alpha_{k}}|\psi_{2}\rangle\langle\psi_{k}|\\
\hline \vdots & \vdots & \ddots & \vdots\\
\hline \sqrt{\alpha_{k}\alpha_{1}}|\psi_{k}\rangle\langle\psi_{1}| &
\sqrt{\alpha_{k}\alpha_{2}}|\psi_{k}\rangle\langle\psi_{2}| & \cdots
& \alpha_{k}|\psi_{k}\rangle\langle\psi_{k}|\end{array}\right)\
,\label{}\end{equation}
 and hence \begin{equation}
\Phi_{2k}^{({\bf
z})}(P_{2k})=\frac{1}{2(k-1)}\left(\begin{array}{c|c|c|c}
(1-\alpha_{1})\mathbb{I}_{2} & -z_{12}M_{12} & \cdots & -z_{1k}M_{1k}\\
\hline -\overline{z}_{12}M_{21} & (1-\alpha_{2})\mathbb{I}_{2} & \cdots & -z_{2k}M_{2k}\\
\hline \vdots & \vdots & \ddots & \vdots\\
\hline -\overline{z}_{1k}M_{k1} & -\overline{z}_{2k}M_{k2} & \cdots & (1-\alpha_{k})\mathbb{I}_{2}\end{array}\right)\ ,\label{}\end{equation}
 where the $2\times2$ matrices $M_{ij}$ are defined as follows \begin{equation}
M_{ij}=\sqrt{\alpha_{i}\alpha_{j}}\,\Big[|\psi_{i}\rangle\langle\psi_{j}|+\sigma_{y}|\overline{\psi_{i}\rangle\langle\psi_{j}}|\sigma_{y}\Big]\ .\label{}\end{equation}

\begin{lemma} \label{Lemma-MM} Matrices $M_{ij}$ satisfy the following
properties:
\begin{enumerate}
\item $M_{ij}M_{ji}=\alpha_{i}\alpha_{j}\,\mathbb{I}_{2}$,
\item $M_{ij}M_{jk}=\alpha_{j}M_{ik}$.
\end{enumerate}
\end{lemma} One proves this lemma by direct calculation. To prove
(\ref{!}) we perform the induction with respect to $k$. For $k=2$
any $\Phi_{4}^{({\bf z})}$ is unitarily equivalent to the Robertson
map $\Phi_{4}$. Suppose now that the theorem is true for $k=n-1$. To prove that it
holds for $k=n$ we use the following well known

\begin{lemma}[Bhatia \cite{Bhatia}] \label{Lemma-B} A block
matrix \[
\left(\begin{array}{cc}
A & X\\
X^{\dagger} & B\end{array}\right)\ ,\]
 with $A\geq0$ and $B>0$, is positive if and only if \begin{equation}
A\geq XB^{-1}X^{\dagger}\ .\label{}\end{equation}
 \end{lemma} Hence \begin{equation}
2(k-1)\, \Phi^{({\bf z})}_{2k}(P_{2k}) \,=\,
\left(\begin{array}{c|c|c|c}
(1-\alpha_{1})\mathbb{I}_{2} & -z_{12}M_{12} & \cdots & -z_{1n}M_{1n}\\
\hline -\overline{z}_{12}M_{21} & (1-\alpha_{2})\mathbb{I}_{2} & \cdots & -z_{2n}M_{2n}\\
\hline \vdots & \vdots & \ddots & \vdots\\
\hline -\overline{z}_{1n}M_{n1} & -\overline{z}_{2n}M_{n2} & \cdots & (1-\alpha_{n})\mathbb{I}_{2}\end{array}\right)\geq0\ ,\label{}\end{equation}
 if and only iff \begin{eqnarray}  \label{>}
 &  & \left(\begin{array}{c|c|c|c}
(1-\alpha_{1})\mathbb{I}_{2} & -z_{12}M_{12} & \cdots & -z_{1,n-1}M_{1,n-1}\\
\hline \overline{z}_{12}M_{21} & (1-\alpha_{2})\mathbb{I}_{2} & \cdots & -z_{2,n-1}M_{2,n-1}\\
\hline \vdots & \vdots & \ddots & \vdots\\
\hline -\overline{z}_{1,n-1}M_{n-1,1} & -\overline{z}_{2,n-1}M_{n-1,2} & \cdots & (1-\alpha_{n-1})\mathbb{I}_{2}\end{array}\right)\geq\nonumber \\
 &  & \frac{\alpha_{n}}{1-\alpha_{n}}\,\left(\begin{array}{c|c|c|c}
\alpha_{1}\mathbb{I}_{2} & z_{1n}\overline{z}_{2n}M_{12} & \cdots & z_{1n}\overline{z}_{n-1,n}M_{1,n-1}\\
\hline \overline{z}_{1n}z_{2n}M_{21} & \alpha_{2}\mathbb{I}_{2} & \cdots & z_{2n}\overline{z}_{n-1,n}M_{2,n-1}\\
\hline \vdots & \vdots & \ddots & \vdots\\
\hline \overline{z}_{1n}z_{n-1,n}M_{n-1,1} &
\overline{z}_{2n}z_{n-1,n}M_{n-1,2} & \cdots &
\alpha_{n-1}\mathbb{I}_{2}\end{array}\right)\ .\end{eqnarray}
 Now let us define a new
set of positive numbers \begin{equation}
\alpha'_{i}:=\frac{\alpha_{i}}{1-\alpha_{n}}\ ,\ \ \ i=1,\ldots,n-1\ ,\label{}\end{equation}
 and new set of matrices $M'_{ij}$ \begin{equation}
M'_{ij}:=\sqrt{\frac{\alpha'_{i}\alpha'_{j}}{\alpha_{i}\alpha_{j}}}\, M_{ij}\ ,\label{}\end{equation}
 for $i,j=1,\ldots,n-1$. It is clear that \begin{equation}
\alpha'_{1}+\ldots+\alpha'_{n-1}=1\ ,\label{}\end{equation}
 and the matrices $M'_{ij}$ satisfy Lemma \ref{Lemma-MM} with $\alpha_{i}$
replaced by $\alpha'_{i}$. Using these new quantities and the
condition $|z_{ij}|\le1$ the inequality (\ref{>}) may be rewritten
as follows \begin{equation} \left(\begin{array}{c|c|c|c}
(1-\alpha'_{1})\mathbb{I}_{2} & -z'_{12}M'_{12} & \cdots & -z'_{1,n-1}M'_{1,n-1}\\
\hline -\overline{z}'_{12}M'_{21} & (1-\alpha'_{2})\mathbb{I}_{2} & \cdots & -z'_{2,n-1}M'_{2,n-1}\\
\hline \vdots & \vdots & \ddots & \vdots\\
\hline -\overline{z}'_{1,n-1}M'_{n-1,1} & -\overline{z}'_{2,n-1}M'_{n-1,2} & \cdots & (1-\alpha'_{n-1})\mathbb{I}_{2}\end{array}\right)\geq0\ ,\label{>>}\end{equation}
 where
\begin{equation}
z'_{ij}:=(1-\alpha_{n})z_{ij}+\alpha_{n}z_{in}\overline{z}_{jn}\ .\label{}\end{equation}
 Note, that \begin{equation}
|z'_{ij}|\leq(1-\alpha_{n})|z_{ij}|+\alpha_{n}|z_{in}\overline{z}_{jn}|
\leq1\ ,\label{}\end{equation}
 due to $|z_{ij}|\leq 1$. Hence inequality (\ref{>>}) is equivalent
to \begin{equation}
\Phi_{2(n-1)}^{({\bf z}')}(P_{2(n-1)})\geq0\ ,\label{!!}\end{equation}
 which is true due to our original assumption that the theorem holds for $k=n-1$. \hfill{}$\Box$

It should be stressed that $\Phi_{2k}^{({\bf z})}$ does not in
general  correspond to the Breuer-Hall map \cite{Breuer,Hall}. One has

\begin{proposition}
A map $\Phi_{2k}^{({\bf z})}$ is equivalent to the Breuer-Hall map iff $z_{ij} =
z_i \overline{z_j}$, where $(z_1,\ldots,z_{2k})$ are defined by $z_k = e^{i\alpha_k}$.
\end{proposition}
Proof: indeed, any such
vector gives rise to the unitary matrix $U^{({\bf z})}$ via
\begin{equation}\label{}
    U^{({\bf z})}_{kl} = \delta_{kl} z_l\ .
\end{equation}
One has

\begin{equation}\label{}
     \Phi_{2k}^{({\bf z})}(X) = U^{({\bf z})}  \Phi_{2k}(X) U^{({\bf z})} \ ,
\end{equation}
and hence $\Phi_{2k}^{({\bf z})}$ is unitary equivalent to the Breuer-Hall map. If $z_{ij} \neq z_i \overline{z_j}$,
then the corresponding entanglement witness $W^{({\bf z})}$ has different spectrum and hence cannot be
equivalent to the entanglement witness corresponding to the Breuer-Hall map.

\begin{proposition} $\Phi_{2k}^{({\bf z})}$, with $|z_{ij}|=1$,
defines an indecomposable map. \end{proposition}

\noindent Proof: let us consider the following  state $\rho$ living
in $\mathbb{C}^{2k} \ot  \mathbb{C}^{2k}$:
\begin{equation}\label{new-states}
\rho^{({\bf z})} =\mathcal{N}\sum_{i,j=1}^{2k} e_{ij}\ot\rho^{({\bf
z})}_{ij} \ ,
\end{equation}
where $\rho^{({\bf z})}_{ij} \in M_{2k}(\mathbb{C})$ are defined as
follows: if $i+j = 2\ell$, then
\begin{equation}\label{}
    \rho^{({\bf z})}_{ij} = - W^{({\bf z})}_{ij}\ .
\end{equation}
If $i+j = 2\ell +1$, one has either
\begin{equation}\label{}
    \rho^{({\bf z})}_{ij} = \mathbb{O}_{2k}\ ,
\end{equation}
for $(i,j) = (2m-1,2m)$ and $m=1,\ldots,k$, or
\begin{equation}\label{}
    \rho^{({\bf z})}_{ij} = \frac{z_{ij}}{4k(k-1)}\ e_{ij}\ ,
\end{equation}
for $(i,j) \neq (2m-1,2m)$. Finally, the normalization constant
reads $ \mathcal{N}=1/3$. One easily checks that $\rho^{({\bf z})}$
defines a PPT state. Now direct calculation shows that
\begin{equation}\label{}
\text{Tr}({W}^{({\bf z})}\rho^{({\bf z})})=-\frac{1}{24k(k-1)}<0\ ,
\end{equation}
which proves that ${W}^{({\bf z})}$ is an indecomposable entanglement witness. \hfill
$\Box$

\begin{corollary}
The formula (\ref{new-states}) defines a new class of PPT entangled
states in $\mathbb{C}^{2k} \ot \mathbb{C}^{2k}$.
\end{corollary}

\subsection{Optimality and SPA}

Finally, let us analyze the problem of optimality of $\Phi^{({\bf
z})}_{2k}$. One has the following

\begin{proposition} $\Phi_{2k}^{({\bf z})}$ is optimal if and only if $|z_{ij}|=1$.
\end{proposition}

\noindent Proof: the necessity of $|z_{ij}|=1$ is obvious (compare
the proof of Proposition \ref{R-opt}). Now, to prove that this
condition is also sufficient we use again the result of Lewenstein
et. al. \cite{O} (cf. Proposition \ref{PRO-Lew}). Let
$z_{kl}=e^{i\alpha_{kl}}$, as before. It is easy to check that the
following vectors \[
f_{kl}=(e_{k}+e^{-i\alpha_{kl}/2}e_{l})\ot(e_{k}+e^{-i\alpha_{kl}/2}e_{l})\
,\ \ \
g_{kl}=(e_{k}+ie^{-i\alpha_{kl}/2}e_{l})\ot(e_{k}-ie^{-i\alpha_{kl}/2}e_{l})\
,\ \ \ e_{k}\ot e_{k}\ ,\]
 span the whole Hilbert space $\mathbb{C}^{2k}\ot\mathbb{C}^{2k}$
and that they satisfy condition: \begin{equation} \<f_{kl}|W^{({\bf
z})}|f_{kl}\>=\<g_{kl}|W^{({\bf z})}|g_{kl}\>=0\ , \ \ \<e_{k}\ot
e_{k}|W^{({\bf z})}|e_{k}\ot e_{k}\>=0.\label{}\end{equation} Thus,
$W^{({\bf z})}=(\oper \ot \Phi_{2k}^{({\bf z})})P^+_{2k}$ is an
optimal entanglement witness. \hfill $\Box$

Concerning SPA we have the following

\begin{proposition} \label{PRO-SPA} SPA for $\Phi_{6}^{({\bf z})}$ and $z_{ij} = -1$
is entanglement breaking.
\end{proposition}

\noindent Proof: consider the following class of states living in
$\Cd \ot \Cd$
\begin{equation}\label{}
    \rho = \sum_{k,l=1}^d a_{ij} e_{ij} \ot e_{ij} + \sum_{i\neq j}
    b_{ij} e_{ii} \ot e_{jj}\ ,
\end{equation}
where the $d \times d$ complex matrix $a_{ij}$ is positive
semidefinite. It was shown \cite{PPT-nasza} that $\rho$ is invariant
under the maximal abelian subgroup of $U(d)$
\begin{equation}\label{UU}
    U_\mathbf{x} \ot \overline{U}_\mathbf{x} \, \rho = \rho\, U_\mathbf{x} \ot \overline{U}_\mathbf{x}\ ,
\end{equation}
where
\begin{equation}\label{}
U_\mathbf{x} = \exp\left(i \sum_{k=0}^{d-1} x_k e_{kk} \right)\ ,
\end{equation}
and $\mathbf{x}=(x_1,\ldots,x_{d}) \in [0,2\pi) \times \ldots \times
[0,2\pi)$. Let $\mathcal{P}$ denotes the following projector
\begin{equation}\label{}
    \mathcal{P}(\rho) := \frac{1}{(2\pi)^d} \int_0^{2\pi} dx_1
    \ldots \int_0^{2\pi} dx_d\  U_\mathbf{x} \ot \overline{U}_\mathbf{x} \,
    \rho (U_\mathbf{x} \ot \overline{U}_\mathbf{x})^\dagger\ ,
\end{equation}
that is, $\mathcal{P}(\rho)$ performs symmetrization of $\rho$ with
respect to $U_{\bf x}$. It is clear that $\mathcal{P}$ maps
separable states into separable states. Now, observe that
\begin{equation}\label{}
    W^{({\bf z})} = \mathcal{P}(V_1) +
    \mathcal{P}[(\oper_{n}\otimes\sigma_{x})V_{2}(\oper_{n}\otimes\sigma_{x})]
    + D\ ,
\end{equation}
where
\begin{equation}\label{}
V_{1} =
\sum_{i=1}^{4}|\psi_{i}\otimes\psi_{i}\rangle\langle\psi_{i}\otimes\psi_{i}|\
, \ \ \ \ V_{2}  =
\sum_{i=1}^{4}|\psi_{i}\otimes\phi_{i}\rangle\langle\psi_{i}\otimes\phi_{i}|\
,
\end{equation}
with
\begin{eqnarray*}
  \psi_{1}  =[1\,0\,1\,0\,1\,0]\ , \ \ \
  \psi_{2}  =[1\,0\,0\,1\,0\,1]\ ,\ \ \
  \psi_{3}  =[0\,1\,1\,0\,0\,1]\ ,\ \ \
  \psi_{4}  =[0\,1\,0\,1\,1\,0]\ ,
\end{eqnarray*}
and
\begin{eqnarray*}
  \phi_{1}  =[1\,0\,1\,0\,1\,0]\ ,\ \ \
  \phi_{2}  =[1\,0\,0\,-1\,0\,-1]\ ,\ \ \
  \phi_{3}  =[0\,1\,-1\,0\,0\,1]\ ,\ \ \
  \phi_{4}  =[0\,1\,0\,1\,-1\,0] \ .\end{eqnarray*}
Finally, $D$ is diagonal. It is clear, that $V_1$ and $V_2$ are
separable. Hence, $W^{({\bf z})}$ is separable being the convex
combination of symmetrized separable operators and diagonal $D$.
\hfill $\Box$

\begin{remark} Clearly the above proposition is trivially  satisfied
for $\Phi_4^{({\bf z})}$ and $z_{12}=-1$. Actually, there is a
strong numerical evidence that SPA for $\Phi_4^{({\bf z})}$ with
$|z_{12}|=1$ is entanglement breaking.
\end{remark}

\section{Conclusions}

We  provided a  generalization of the well known linear positive maps:
reduction map in $\Mn$ and Robertson map in $M_{2k}(\mathbb{C})$:
$R^{({\bf z})}_n$ and  $\Phi^{({\bf z})}_{2k}$, respectively. We
showed that for each collection $z_{ij}$ ($i < j$) satisfying
$|z_{ij}|\leq 1$ these maps are positive. Hence, each collection of points
from the unit disc in the complex plane $\mathbb{C}$ gives rise to
a positive map. Interestingly, points from the boundary, i.e. satisfying
 $|z_{ij}|=1$, generate optimal maps: decomposable in the case of reduction map
and indecomposable in the case of Robertson map.

Our construction gives rise to the new classes of entanglement
witnesses: decomposable entanglement witnesses corresponding to
$R^{({\bf z})}_n$, and indecomposable entanglement witnesses
corresponding to $\Phi^{({\bf z})}_{2k}$. As a byproduct we provided
new examples of PPT entangled states in $\mathbb{C}^{2k} \ot
\mathbb{C}^{2k}$ detected by indecomposable entanglement witnesses.
Our analysis supports recent conjecture \cite{SPA3,SPA4} that
structural physical approximation to an optimal positive map defines
entanglement breaking completely positive map. Actually, we were
able to prove it for generalized reduction map. Concerning
generalized Robertson map Proposition~\ref{PRO-SPA} provides
evidence that it supports conjecture \cite{SPA3,SPA4} as well.

\section*{Acknowledgement}

This work was partially supported by the Nicolaus Copernicus Grant
383-F.


\bibliographystyle{plain}

\end{document}